\documentclass[aps,pra,reprint,superscriptaddress]{revtex4-2}

\usepackage{graphicx}
\usepackage{amsmath}
\usepackage{amsfonts}
\usepackage{color}

\DeclareMathOperator{\Tr}{Tr}

\begin{document}

\title{Laughlin topology on fractal lattices without area law entanglement}

\author{Xikun Li}
\thanks{These authors contributed equally to this manuscript.}
\affiliation{School of Physics and Optoelectronics Engineering, Anhui University, Hefei, Anhui 230601, China}
\affiliation{Max-Planck-Institut f\"ur Physik komplexer Systeme, D-01187 Dresden, Germany}

\author{Mani Chandra Jha}
\thanks{These authors contributed equally to this manuscript.}
\affiliation{Max-Planck-Institut f\"ur Physik komplexer Systeme, D-01187 Dresden, Germany}

\author{Anne E. B. Nielsen}
\affiliation{Max-Planck-Institut f\"ur Physik komplexer Systeme, D-01187 Dresden, Germany}
\affiliation{Department of Physics and Astronomy, Aarhus University, DK-8000 Aarhus C, Denmark}

\begin{abstract}
Laughlin states have recently been constructed on fractal lattices, and the charge and braiding statistics of the quasiholes were used to confirm that these states have Laughlin type topology. Here, we investigate density, correlation, and entanglement properties of the states on a fractal lattice derived from a Sierpinski triangle with the purpose of identifying similarities and differences compared to two-dimensional systems and with the purpose of investigating whether various probes of topology work for fractal lattices. Similarly to two-dimensional systems, we find that the connected particle-particle correlation function decays roughly exponentially with the distance between the lattice sites measured in the two-dimensional plane, but the values also depend on the local environment. Contrary to two-dimensional systems, we find that the entanglement entropy does not follow the area law if one defines the area to be the number of nearest neighbor bonds that cross the edge of the selected subsystem. Considering bipartitions with two bonds crossing the edge, we find a close to logarithmic scaling of the entanglement entropy with the number of sites in the subsystem. This also means that the topological entanglement entropy cannot be extracted using the Kitaev-Preskill or the Levin-Wen methods. Studying the entanglement spectrum for different bipartitions, we find that the number of states below the entanglement gap is robust and the same as for Laughlin states on two-dimensional lattices.
\end{abstract}

\maketitle

\section{Introduction}

Entanglement plays an important role in gaining insights into the physics of strongly correlated quantum many-body systems. The entanglement entropy of the ground state of local, gapped Hamiltonians, typically follows an area law, which means that the entanglement entropy depends linearly on the size of the boundary of the considered subsystem \cite{eisert2010}. Topological systems are further characterized by the topological entanglement entropy, which appears as a constant and universal term in addition to the linear term in the area law \cite{kitaev2006,levin2006,jiang2012identifying}. The entanglement spectrum contains more detailed information about the topology \cite{li2008entanglement,lauchli2010, dubail2012real,sterdyniak2012,dubail2012edge,regnault2017entanglement}.

Recently, interest has developed with respect to investigating topological phases of matter on fractal lattices \cite{brzezinska2018topology,pai2019topological,manna2019,iliasov2020, fremling2020existence,manna2021laughlin,sarangi2021,guanyu2021,fremling2021, ivaki2021topological}, such as the Sierpinski triangle. This is, in part, motivated by experimental developments, e.g., within molecular \cite{shang2015assembling} and electronic \cite{kempkes2019design} fractals. Fractals are self-similar structures that look the same when zooming in or out. In fractal lattices with fractional Hausdorff dimension, there is no clear distinction between bulk and edge, and they also lack the periodicity present in traditional lattices. These unusual properties, coupled with the observation that topological order can be found in unconventional (e.g., amorphous or quasiperiodic) lattices has naturally led to curiosity about the nature of topological phases on fractal lattices.

Topological phases of non-interacting systems, including integer quantum Hall phases, have been found on fractal lattices \cite{brzezinska2018topology,pai2019topological,iliasov2020, fremling2020existence,sarangi2021,fremling2021,ivaki2021topological}. Recently, fractional quantum Hall models have also been constructed and investigated on such lattices \cite{manna2019,manna2021laughlin}. It was shown in \cite{manna2019} that one can define a trial state on fractal lattices that is closely related to the Laughlin state with $q$ fluxes per particle, where $q$ is a positive integer. By computing the charge and braiding properties of quasiholes inserted into the state, it was shown for $q=2$ and $q=3$ that this state has the same topology as the original Laughlin state. This opens several questions. In particular, how do the properties of topologically ordered states on a fractal lattice compare to the properties on two-dimensional lattices, and how can one detect topological properties on fractal lattices? The latter is nontrivial as many of the measures that are commonly used for two-dimensional systems rely on assumptions that are not necessarily true for fractal lattices. The fractal lattices do, e.g., not have periodic boundary conditions, and hence measures such as Chern number and spectral flow do not translate straightforwardly to fractal lattices.

Here, we take a closer look at the properties of Laughlin states on fractal lattices compared to Laughlin states on two-dimensional lattices, and we investigate the possibilities to use entanglement measures to detect the topology in the states. We first show that the particle density has more structure on the fractal lattice than for two-dimensional systems, and that the particle-particle correlation function decays roughly exponentially with distance, but also depends on the local lattice structure. We then investigate the entanglement entropy. We find that the dependence of the entanglement entropy on the number of particles in the state shows oscillations that are not present for two-dimensional square lattices. We also find that the area law is generally not fulfilled on the fractal lattice, if we define the size of the boundary to be the number of nearest neighbor bonds that cross the edge of the selected subsystem. This means that we cannot extract the topological entanglement entropy using the Kitaev-Preskill or the Levin-Wen methods. The entanglement spectrum, in contrast, has the same structure as for the two-dimensional systems and can hence provide information about the topology.

The article is structured as follows. In Sec.\ \ref{sec:states}, we introduce the fractal lattice and the Laughlin states that we consider in this article. In Sec.\ \ref{sec:density}, we compare the particle density for fractal and two-dimensional lattices. In Sec.\ \ref{sec:cor}, we show that the particle-particle correlations decay roughly exponentially with distance. In Sec.\ \ref{sec:entropy}, we study the entanglement entropy of the states. We first establish that the entropy converges both in the infinite generation limit and in the thermodynamic limit. We then quantify how the entropy depends on the number of particles in the system and the size of the subsystem. We also show that the methods to compute the topological entanglement entropy proposed by Kitaev and Preskill and by Levin and Wen fail. In Section \ref{sec:ES}, we compute the entanglement spectrum and show that the number of states below the entanglement gap is the same as for Laughlin states on two-dimensional lattices. Section \ref{sec:conclusion} summarizes the conclusions.

\section{Laughlin states on fractal lattices}\label{sec:states}

\begin{figure}
\includegraphics[width=0.75\linewidth,trim=134 264 117 254,clip]{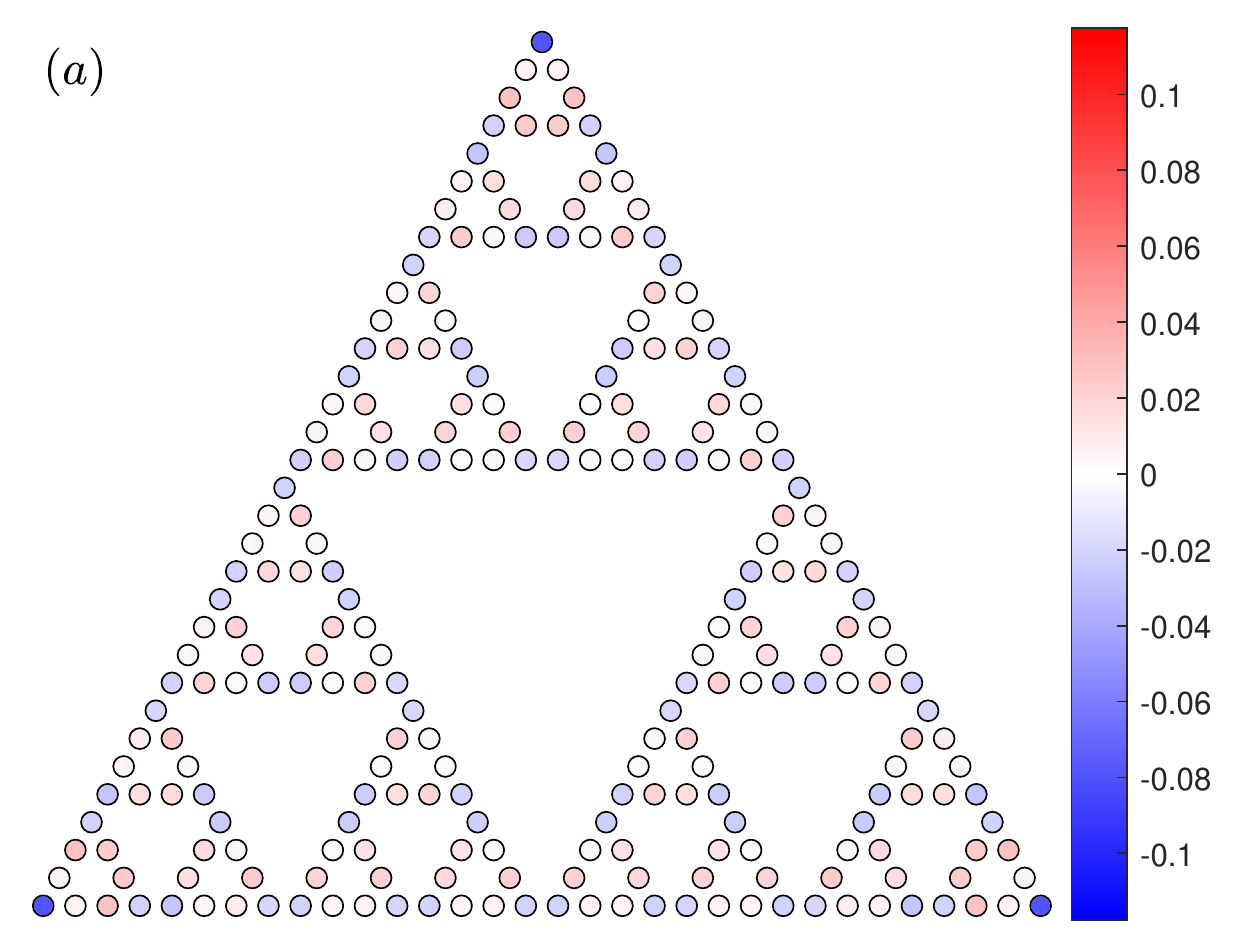}
\includegraphics[width=0.75\linewidth,trim=134 264 117 254,clip]{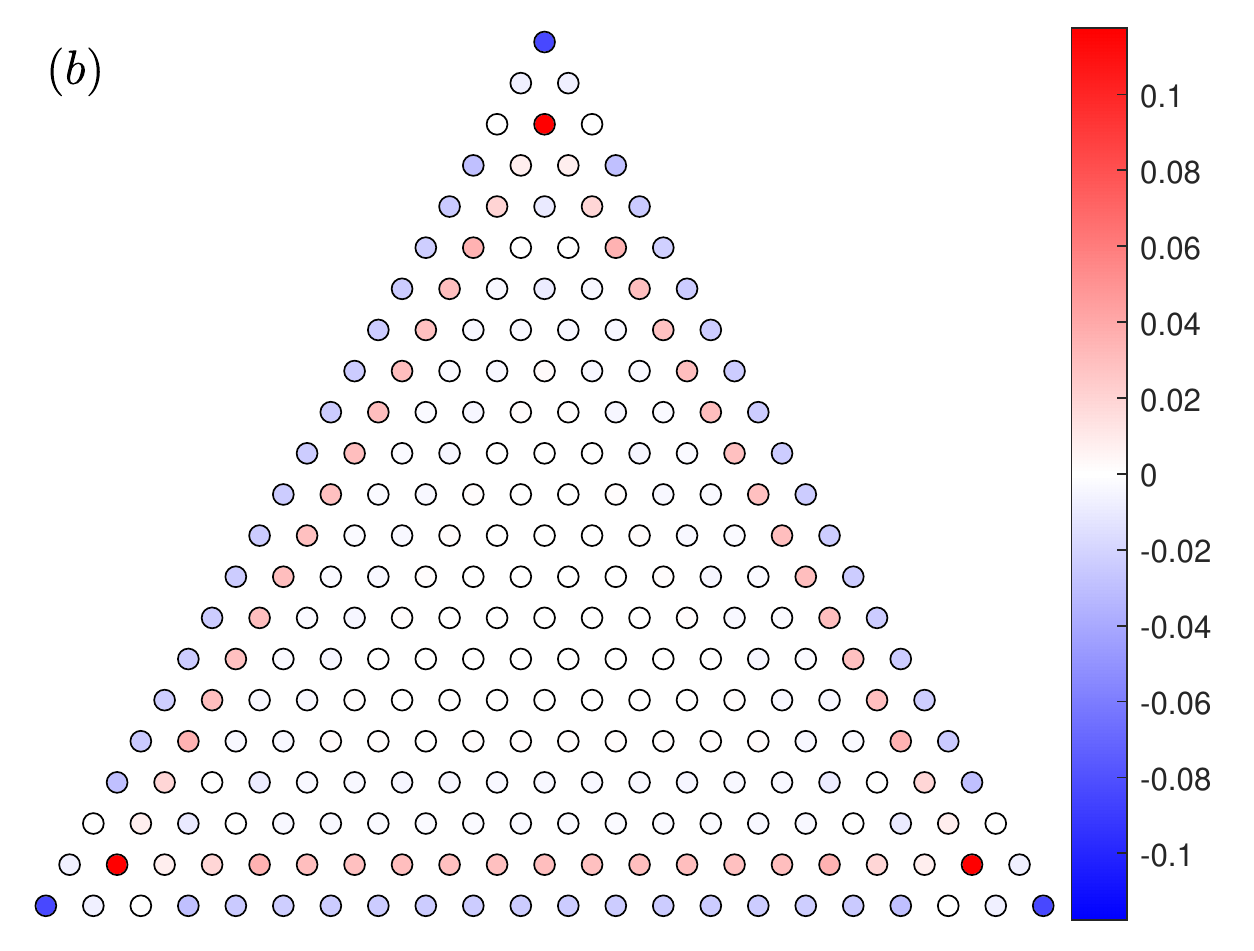}
\caption{(a) The fractal lattice considered in this article (here shown for generation $g=5$). The color of the $j$th site shows $\langle n_j \rangle-M/N$, where $\langle n_j \rangle$ is the expectation value of the number of particles on the $j$th site for the Laughlin state \eqref{laughlin} computed from Monte Carlo simulations and $M/N$ is the average number of particles on a site. The lattice has $N=243$ sites, there are $M=27$ particles in the system, and the number of flux units per particle is $q=3$. (b) The same, but for a triangular lattice with $253$ sites and $28$ particles.}
\label{fig:density}
\end{figure}

We consider the lattice illustrated in Fig.~\ref{fig:density}(a), which is obtained by putting one lattice site at the center of each of the triangles making up a Sierpinski triangle. The lattice consists of $N=3^g$ sites, where $g$ is the generation of the fractal lattice. In physical systems, the generation is always finite. Here, we are primarily interested in a regime, where there are several particles in the system, but several lattice sites available for each particle. We shall therefore typically take $1\ll M\ll N$, where $M$ is the number of particles.

We consider Laughlin type states on this lattice. Such states have the property that the wavefunction amplitude goes to zero, when two particles approach each other, and the particles therefore tend to stay away from each other. When we increase the generation of the fractal lattice by one, we effectively split each lattice site into three lattice sites. If the distance between neighboring lattice sites is already much smaller than the typical distance between the particles, however, this will not change the physics significantly. It does also not make a significant difference, whether the sites are points or small triangles, since there will anyway be at most one particle on each triangle. As pointed out also in \cite{manna2019}, this means that if we are in the regime where $1\ll M\ll N$, the physics is effectively the physics of the state defined on the infinite generation Sierpinski triangle, which has Hausdorff dimension $D=\ln(3)/\ln(2)\approx1.585$. We provide an example of this convergence in Sec.\ \ref{sec:infgen} below.

Following \cite{manna2019}, we define the Laughlin type trial states
\begin{align}\label{laughlin}
&|\psi\rangle=\sum_{n_1,\ldots,n_N} \psi(n_1,\ldots,n_N)
|n_1,\ldots,n_N\rangle,\\
&\psi(n_1,\ldots,n_N) \propto
\delta_n \, \prod_{i<j} (z_i-z_j)^{q n_i n_j}
\prod_{k\neq l} (z_k-z_l)^{-\eta n_l},\nonumber
\end{align}
where $z_j$ is the position of the $j$th site in the two-dimensional plane written as a complex number, $n_j\in\{0,1\}$ is the number of particles on the $j$th site, $q$ is a positive integer giving the number of flux lines per particle, and the delta function $\delta_n$ is one for terms with $M\equiv \sum_j n_j=\eta N/q$ particles and zero otherwise. The particles are fermions for $q$ odd and hardcore bosons for $q$ even.

The exponent $-\eta n_l$ appearing in \eqref{laughlin} is not necessarily an integer. Different choices of the roots lead to wavefunctions that differ by a simple, unitary transformation. All the properties we compute below are, however, independent of the choice. The density and particle-particle correlations only depend on the norm of the wavefunction, the factors involving $\eta$ cancel in the fraction appearing in the expression \eqref{replica} for the entropy, and the Schmidt coefficients in \eqref{schmidt} from which the entanglement spectrum is obtained are also not affected. We shall hence not make a particular choice here.

The states in \eqref{laughlin} are unchanged if all $z_i$ are multiplied by the same factor, which corresponds to scaling and/or rotating the lattice in the complex plane. This result can be derived by utilizing that $n_i^2=n_i$ and that $\sum_i n_i$ is a constant. We also note that renumbering the sites leaves the state unchanged for $q$ even, while for $q$ odd, renumbering the sites corresponds to choosing a different ordering of the fermion creation operators in the basis states. The numbering can hence be chosen after convenience.

The states in \eqref{laughlin} differ from the Laughlin states in the two-dimensional plane in two ways. First, the particles are only allowed to be on the lattice sites, rather than anywhere in the plane. Second, the uniform magnetic field in the two-dimensional plane is replaced by a magnetic field going through the lattice sites only. This is natural, since the latter corresponds to a uniform magnetic field on the fractal lattice. It follows from the relation $M=\eta N/q$ that $\eta N$ is the total magnetic flux. The parameter $\eta$ is hence the magnetic flux through one site. We note that exact parent Hamiltonians for the states in \eqref{laughlin} are provided in Ref.~\cite{tu2014}.

It has already been demonstrated for $q=2$ and $q=3$ in \cite{manna2019} that the states \eqref{laughlin} defined on the fractal lattice illustrated in Fig.\ \ref{fig:density}(a) have the same topological properties as the Laughlin states in the two-dimensional plane. This was done by showing that one can insert quasiholes, which have the same fractional charge and braiding properties as quasiholes in the Laughlin states. In the following sections, we investigate several further properties of the states and find both similarities and differences compared to the properties of Laughlin states on two-dimensional lattices.

\begin{figure*}
\includegraphics[width=\linewidth,trim=132 490 133 122,clip]{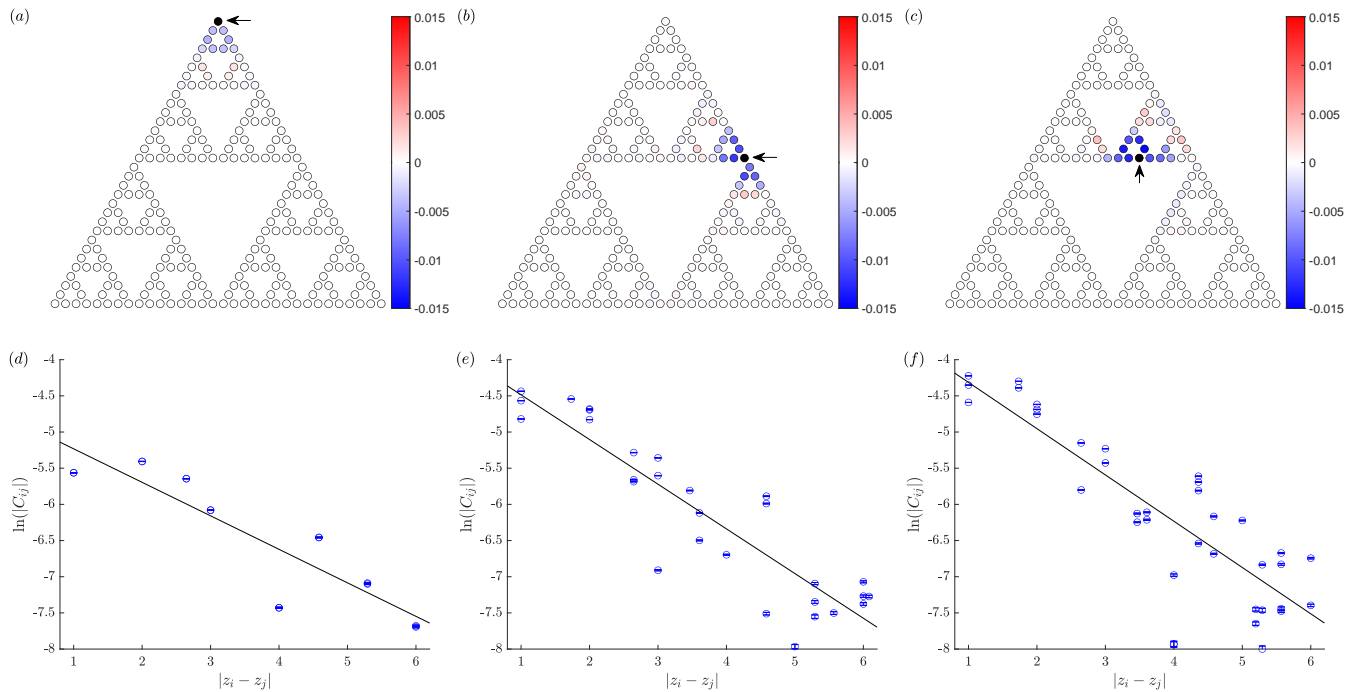}
\caption{(a-c) The particle-particle correlation function $C_{ij}$ in \eqref{cor} between a fixed lattice site $i$ and all other lattice sites $j$ for $N=243$, $M=27$, and $q=3$. The site $i$ is shown in black and is also indicated with an arrow. The value of $C_{ij}$ is shown with color for all $j\neq i$. (d-f) Logarithm of the absolute value of the correlation function $\ln(|C_{ij}|)$ versus the distance $|z_i-z_j|$ for the three choices of $i$ considered above. We here measure distances in units of the distance between nearest neighbor sites. The blue circles with error bars are the values computed from Monte Carlo sampling, while the black solid line is a fit of the form $a-|z_i-z_j|/\xi$ to the shown data points, where the constant $a$ and the correlation length $\xi$ are the fitting parameters. The correlation lengths extracted from the fits are (d) $\xi=2.16$, (e) $\xi=1.62$, and (f) $\xi=1.56$, respectively.}
\label{fig:corr}
\end{figure*}

\section{Density}\label{sec:density}

We first consider the particle density. The particle density $\langle n_i \rangle$ of the Laughlin states on two-dimensional lattices is constant in the bulk and show oscillations near the edge \cite{rodrigues}. Figure \ref{fig:density} shows the deviation of the particle density from the average density for the fractal lattice and for comparison also for a triangular lattice with a triangular edge. We have obtained these results through Metropolis Monte Carlo simulations of
\begin{equation}
\langle n_i\rangle=\sum_{n_1,\ldots,n_N} n_i |\psi(n_1,\ldots,n_N)|^2.
\end{equation}
Oscillations in the density are seen on the three edges of the triangular lattice, while the density takes the average value $M/N$ in the bulk. An important difference between the triangular lattice and the fractal lattice is that the fractal lattice has edges everywhere. This leads to a more complicated density pattern. The three corner sites have a density of $0.0327$, which is significantly below the average $M/N=1/9$, but the density on all other sites varies at most 25\% from the average value. We observe similar density patterns for larger generations, e.g.\ $g=6$, and for different values of $M$ and $q$.

\section{Correlations}\label{sec:cor}

To quantify over how long distances the particles sense each other, we next compute the particle-particle correlation function
\begin{equation}\label{cor}
C_{ij}= \langle n_i n_j\rangle-\langle n_i \rangle \langle n_j\rangle
\end{equation}
through Metropolis Monte Carlo simulations of
\begin{equation}
\langle n_in_j\rangle=\sum_{n_1,\ldots,n_N} n_in_j |\psi(n_1,\ldots,n_N)|^2.
\end{equation}
For Laughlin states on two-dimensional lattices, the particle-particle correlation function decays exponentially with a correlation length of order $1$ \cite{glasser2016}. Results for the fractal lattice with $N=243$, $M=27$, and $q=3$ are shown in Fig.~\ref{fig:corr}. The particle-particle correlations are seen to roughly decay exponentially with distance, although there is also some dependence on the local lattice structure. For correlations with the corner site, the correlation length obtained from the plot is about $2.2$, while it is about $1.6$ for the sites that are not at the corners. We also note that the correlations are negative for short distances, which expresses the repulsion between the particles in the Laughlin states.

\section{Entanglement entropy}\label{sec:entropy}

The entanglement entropy is another important quantity to characterize the behavior of quantum many-body systems. Here, we study the R\'{e}nyi entropy with index two, since it can be computed with Monte Carlo simulations using the replica trick \cite{ciracsierra,melko2010}. It is defined as $S_A=-\ln\left[\Tr\left(\rho_A^2\right)\right]$, where $\rho_A=\mathrm{Tr}_B(|\psi\rangle\langle\psi|)$ is the reduced density operator of a subsystem $A$ containing $N_A$ sites, and $B$ is the part of the system not belonging to $A$. The replica trick is to note that
\begin{multline}\label{replica}
e^{-S_A}=\\
\sum_{n,m,n',m'}
\frac{\psi(n,m')\psi(n',m)}{\psi(n,m)\psi(n',m')}
|\psi(n,m)|^2 |\psi(n',m')|^2,
\end{multline}
and compute the right hand side using Monte Carlo simulations. Here, $n=\{n_1,n_2,\ldots,n_{N_A}\}$ is a basis for subsystem $A$, $m=\{n_{N_A+1},n_{N_A+2},\ldots,n_{N}\}$ is a basis for subsystem $B$, and $n'$ and $m'$ are independent copies of $n$ and $m$. We always choose the numbering of the sites such that the sites in $A$ are numbered from $1$ to $N_A$, and the sites in $B$ are numbered from $N_A+1$ to $N$. This is important for fermions, but does not make a difference for bosons.

The entanglement entropy of Laughlin states on two-dimensional lattices satisfies an area law, $S_A=\alpha L-\gamma$, where $L$ is the length of the boundary, $\alpha$ is a nonuniversal constant, and $\gamma$ is the topological entanglement entropy, which is a universal constant providing information about the topology of the system \cite{kitaev2006,levin2006}. For Laughlin states the topological entanglement entropy is $\gamma=\ln(q)/2$. Kitaev and Preskill~\cite{kitaev2006} and Levin and Wen~\cite{levin2006} proposed two similar methods to extract the value of $\gamma$ by dividing the system into different regions and adding and subtracting entropies of these regions in a particular pattern. It has also been observed in \cite{glasser2016}, that the entropy has the symmetry $S_A(M)=S_A(N-M)$, i.e., replacing empty sites with filled sites and filled sites with empty sites does not change the entropy.

Below, we first show that the entropy of the states on the fractal lattice converges in the infinite generation limit and in the thermodynamic limit. We then study the dependence of the entropy on the particle number, observing the symmetry mentioned above, but also oscillations that do not occur for two-dimensional square lattices. Investigating the dependence of the entropy on the subsystem size, we find that the system does not follow the area law. Instead, if we choose a subsystem whose boundary cuts two nearest neighbor bonds, we find a close to logarithmic increase in the entropy with the number of sites in the subsystem. This result means that we are not able to obtain the topological entanglement entropy from the Kitaev-Preskill or Levin-Wen method. We nevertheless do the computation and find results that are significantly below the topological entanglement entropy for Laughlin states.

\subsection{Infinite generation limit}\label{sec:infgen}

\begin{figure}
\includegraphics[width=\linewidth,trim=5mm 4mm 5mm 4mm]{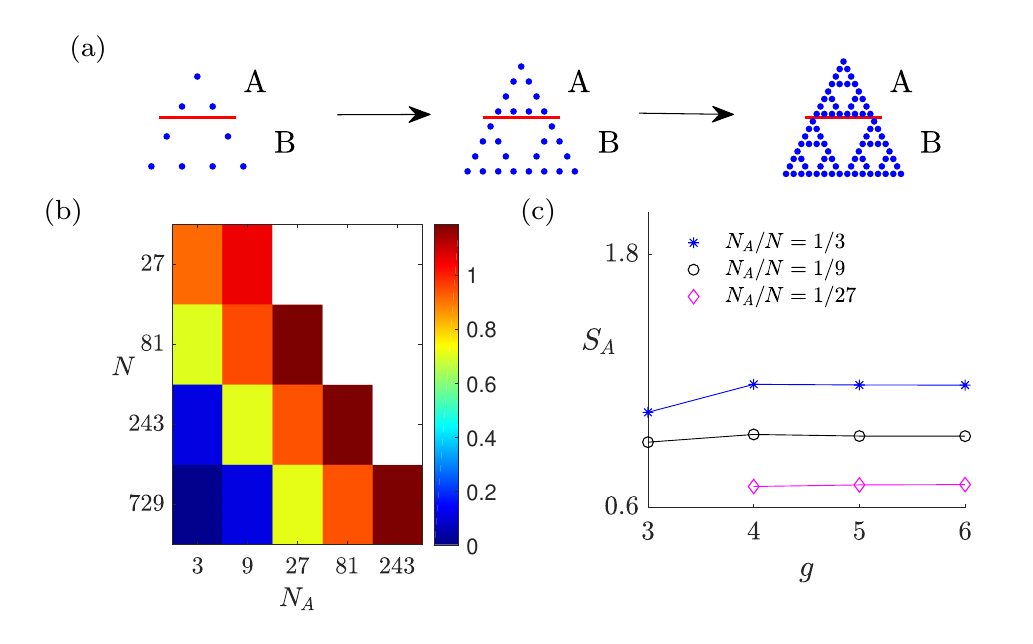}
\caption{Convergence of the entropy $S_A$ in the infinite generation limit. (a) We consider a bipartition, in which subsystem $A$ (the sites above the horizontal line) consists of $N_A=3^m$ sites, where $m$ is a positive integer. (b) $S_A$ as a function of $N$ and $N_A$ and (c) $S_A$ as a function of $g$ for $q=3$ and $M=12$. For fixed $M$ and fixed $N_A/N$, $S_A$ is seen to converge as the generation $g$ increases, i.e., increasing the amount of detail of the lattice does not change the entropy.}
\label{fig:infinite}
\end{figure}

To show that the physics does not depend on whether the generation of the fractal lattice is finite or infinite as long as the generation is large enough, we calculate the entropy for a region containing a fixed fraction $N_A/N$ of the lattice as shown in Fig.~\ref{fig:infinite}(a). We keep $M$ fixed and vary the generation $g$, which corresponds to adding further and further detail to the lattice. From the results in Fig.~\ref{fig:infinite}(b-c), which are computed for $M=12$, $q=3$, and $g\in\{3,4,5,6\}$, we observe that the R\'{e}nyi entropy $S_A$ converges as the generation increases. The same conclusions are obtained for different values of $q$ and $M$. Hence, when the generation is large enough, it does not make a significant difference whether the generation is increased even further.

\subsection{Thermodynamic limit}\label{sec:therm_lim}

\begin{figure}
\includegraphics[width=\linewidth,trim=5mm 5mm 4mm 3mm]{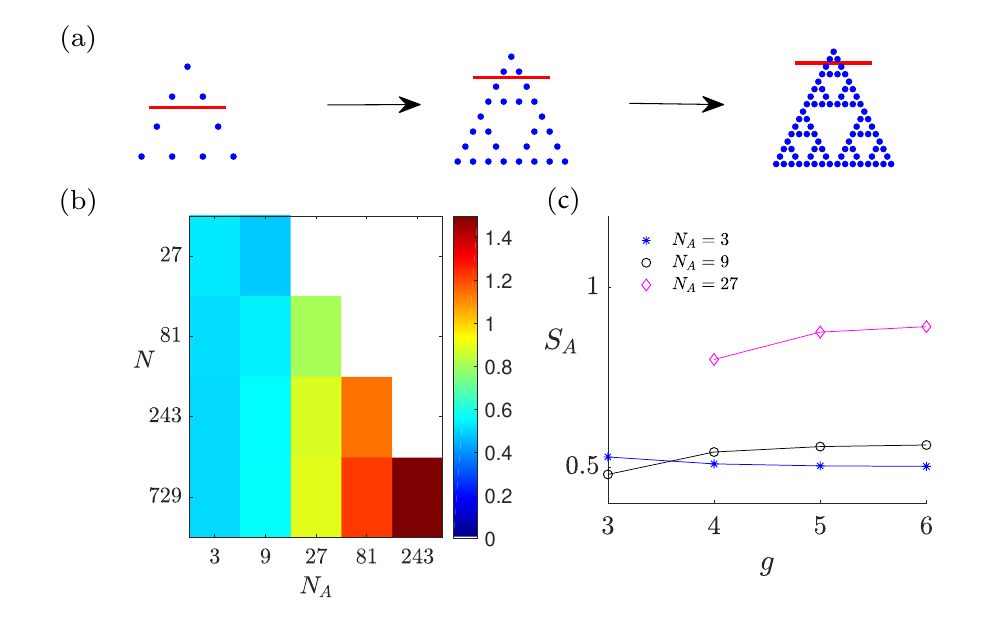}
\caption{Convergence of the entropy $S_A$ in the thermodynamic limit. (a) We consider a bipartition, in which subsystem $A$ (the sites above the horizontal line) consists of a fixed number of sites $N_A$. We also take the ratio $M/N$ between the number of particles and the number of lattice sites to be constant. (b) $S_A$ as a function of $N$ and $N_A$ and (c) $S_A$ as a function of $g$ for $M/N=1/9$ and $q=3$. For fixed $M/N$ and fixed $N_A$, it is seen that $S_A$ converges in the limit of large $N=3^g$, i.e., the entropy of a local region is not affected by how large the lattice is.}
\label{fig:thermo}
\end{figure}

We next show that the entropy also converges in the thermodynamic limit. For this purpose, we consider a region with a fixed number of sites $N_A$ as shown in Fig.~\ref{fig:thermo}(a), and we keep the number of particles per lattice site $M/N$ fixed as we increase the generation of the lattice.
Results for $M/N=1/9$ and $q=3$ are shown in Fig.~\ref{fig:thermo}(b-c), and it is seen that $S_A$ converges as the thermodynamic limit is approached by increasing the generation $g$. We also find convergence of the entropy in the thermodynamic limit for other values of $N_A$, $M/N$, and $q$.

\subsection{Dependence on particle number}

\begin{figure}
\includegraphics[width=0.75\linewidth,trim=30mm 20mm 40mm 28mm,clip]{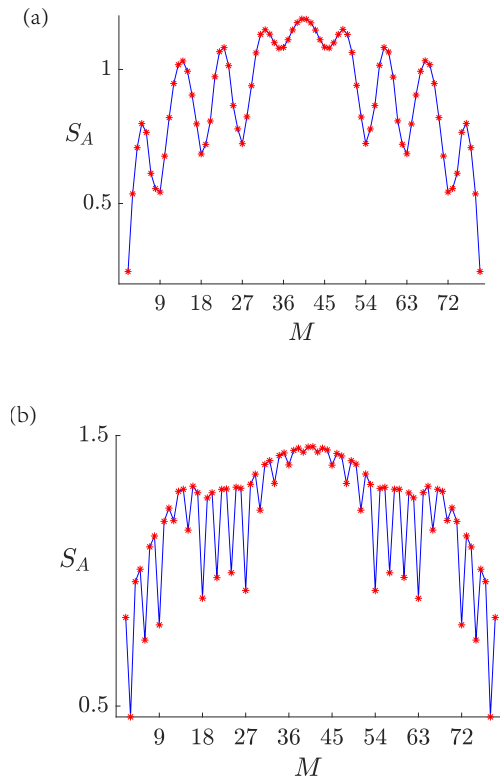}
\caption{R\'{e}nyi entropy $S_A$ versus particle number $M$ for a fractal lattice with $N=81$ sites and $q=3$. In (a), the subsystem $A$ is the top 9 sites of the fractal lattice, and in (b), the subsystem $A$ is the top 27 sites. In (a), local minima are observed for $M=9n$, where $n$ is a positive integer, while in (b), local minima are observed for $M=3n$. In addition, we observe the symmetry $S_A(M)=S_A(N-M)$.}
\label{fig:SM}
\end{figure}

The entropy $S_A$ as a function of the number of particles $M$ for $N=81$, $q=3$, and two different bipartitions is shown in Fig.~\ref{fig:SM}. In Ref.~\cite{glasser2016}, it was noted that the Laughlin states defined on two-dimensional square lattices have the symmetry that the entropy is the same for $M$ particles and for $N-M$ particles, i.e.\ $S_A(M)=S_A(N-M)$. Figure \ref{fig:SM} shows that this symmetry is also present for the fractal lattice.

While the entropy varies quite smoothly with $M$ for the $q=4$ state on a two-dimensional square lattice on a cylinder (see Fig.\ 3 in \cite{glasser2016}), we observe, yet unexplained, oscillations in Fig.\ \ref{fig:SM} with period $\delta M=9$ for the subsystem with $9$ sites and period $\delta M=3$ for the subsystem with $27$ sites. Thus for the results shown in Fig.~\ref{fig:SM}, the product of $N_A$ and the spacing between local minima $\delta M$ is always the total number of sites: $N_A\times \delta M =N$. Furthermore, when we remove one or two sites from the subsystem $A$ and do the same calculations, the oscillation behavior of $S_A$ against $M$ still exists. We obtain similar results for $q=2$ and for larger generations.

\subsection{Dependence on subsystem size}\label{sec:layercut}

\begin{figure}
\includegraphics[width=\linewidth,trim=5mm 5mm 7mm 4mm]{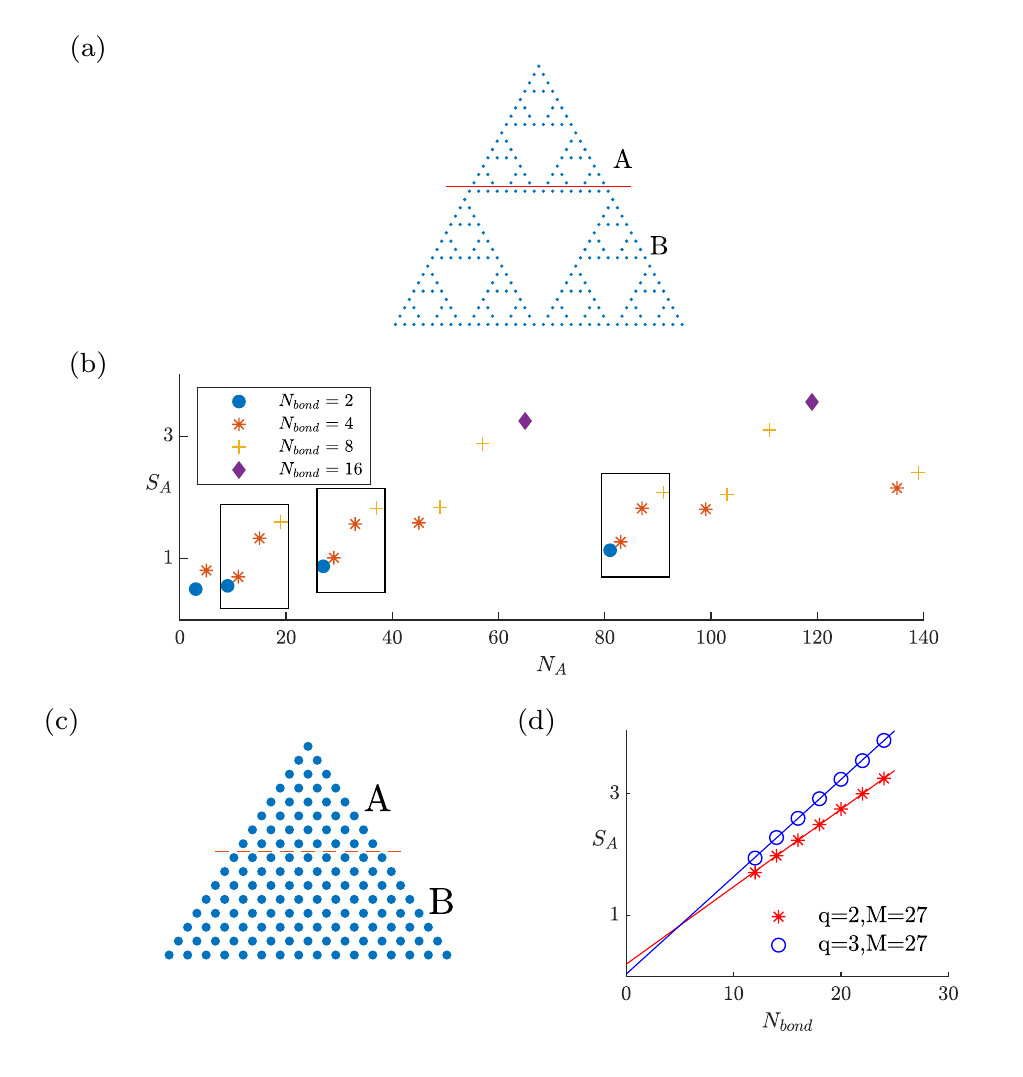}
\caption{(a) Illustration of the layer cut bipartition. Subsystem $A$ consists of the sites above the red line, and subsystem $B$ consists of the sites below the red line. The red line is always horizontal and is moved downwards layer by layer. (b) R\'{e}nyi entanglement entropy $S_A$ as a function of the number of sites $N_A$ in subsystem $A$ using the layer cut bipartition with $q=3$, $N=243$, and $M=27$. The value of $N_{\mathrm{bond}}$ is indicated with different types of markers. (c) We also consider the layer cut bipartition for a triangular lattice with a triangular edge. (d) R\'{e}nyi entanglement entropy $S_A$ for the triangular lattice with $N=136$ sites and $M=27$ particles. The lines are linear least squares fits of the form $S_A=\alpha N_{\mathrm{bond}}-\gamma'$, where the fitting parameters are $\alpha=0.1275$ and $\gamma'=-0.1965$ for $q=2$ and
$\alpha=0.1602$ and $\gamma'=-0.03423$ for $q=3$. Although the data points fall on straight lines, the topological entanglement entropy cannot be extracted from the fits, because the presence of the edges lead to corrections to the term $\alpha N_{\mathrm{bond}}$ that are of the same order as $\gamma$.}
\label{fig:layer_tri}
\end{figure}

To explore the entanglement structure further, we next divide the system as shown in Fig.\ \ref{fig:layer_tri}(a). Here, we cut the system by a horizontal line and move this line vertically down layer-by-layer such that subsystem $A$ increases layer-by-layer. We call this bipartition for layer cut bipartition. Let $N_{\mathrm{bond}}$ denote the number of nearest-neighbor bonds that cross the boundary of the selected subsystem. For the layer cut bipartition, the value of $N_{\mathrm{bond}}$ varies from 2 to $2^g$, and different subsystems $A$ can have the same value of $N_{\mathrm{bond}}$. For the chosen division, $N_{\mathrm{bond}}$ is equal to the number of sites in the layer just below the boundary.

\begin{figure}
\includegraphics[width=0.7\linewidth,trim=40mm 85mm 40mm 88mm,clip]{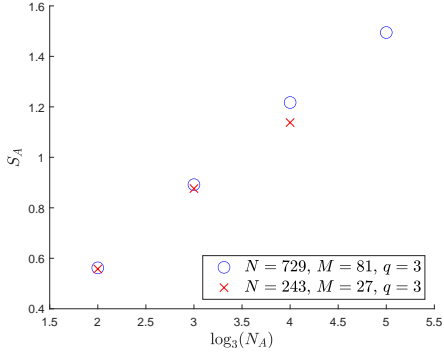}
\caption{The entropy $S_A$ for two fractal lattices with $q=3$ and the same number of particles per site. The region $A$ is chosen to consist of $N_A=3^{m}$ sites from the top of the fractal, where $m$ is taken to be an integer. For this choice, the boundary of subsystem $A$ always cuts two nearest-neighbor bonds. The entropy is seen to increase slightly slower than linear with $\log_3(N_A)$. The curve for $N=729$ is straighter than the one for $N=243$, so the deviation from a linear scaling may be due to the finite system size. This shows that the area law does not apply, as the area law predicts $S_A$ to be independent of $\log_3(N_A)$.}
\label{fig:scale}
\end{figure}

The entropy $S_A$ as a function of the number of sites $N_A$ in subsystem $A$ is shown in Fig.\ \ref{fig:layer_tri}(b), where different values of $N_{\mathrm{bond}}$ are indicated with different types of markers. For comparison, we also show results for $S_A$ for the triangular lattice using the layer cut bipartition in Fig.\ \ref{fig:layer_tri}(c-d). While $S_A$ depends linearly on $N_{\mathrm{bond}}$ for the triangular lattice, this is not the case for the fractal. For the fractal, $S_A$ may take quite different values for different data points corresponding to the same value of $N_{\mathrm{bond}}$. Looking at Fig.\ \ref{fig:layer_tri}(b), we see a certain sequence of data points that restarts at every blue dot (corresponding to $N_{\mathrm{bond}}=2$) and becomes longer and longer as $N_A$ increases. This is a result of the self-similar nature of the fractal lattice, and it leads to the repetitions of patterns, such as those inside the rectangles. Considering the two data points with $N_{\mathrm{bond}}=4$ inside the leftmost rectangle, it is not surprising that the entropy is not the same for the two data points, as the distribution of lattice sites close to the boundary is quite different in the two cases. If the area law applied, however, we would expect points at the same position in the repeating sequences to have the same entropy, if $N_A$ is not too small. The blue dots (corresponding to $N_{\mathrm{bond}}=2$) is such a set of points, and we plot these data points as a function of $\log_3(N_A)$ in Fig.\ \ref{fig:scale} together with the corresponding data points for a lattice with $729$ sites. We observe that $S_A$ increases approximately linearly with $\log_3(N_A)$ instead of being constant as predicted by the area law. The area law is hence not followed. The curves bend slightly, but the bending is less for the larger system, so the bending could be due to the finite size of the considered lattices.

Taking a closer look at Fig.\ \ref{fig:layer_tri}(b), we observe that local minima of $S_A$ occur at $N_{A}\in\{3,9,27,45,81,99,135\}$. At these values, the boundary lies directly above certain numbers of blocks of 9 sites. The local maxima lie at $N_{A}\in\{5,19,37,65,91,119\}$, which correspond to one layer above the divisions that produce the minima of $S_A$. To investigate the origin of the maxima and minima, we check the robustness of these peaks and troughs by computing $S_A$ for different parameters. While the value of $S_A$ at the peaks and troughs varies, the position of the maxima and minima are robust for a range of different choices of $M$, $N$, and $q$. Therefore, this structure is due to the intrinsic properties of the fractal lattice.

\subsection{Topological entanglement entropy}\label{sec:TEE}

\begin{figure}
\includegraphics[width=\linewidth,trim=22mm 124mm 25mm 18mm,clip]{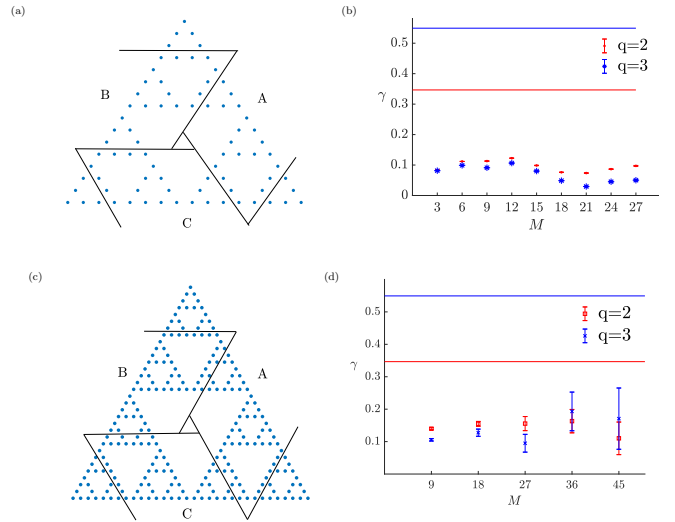}
\caption{(a) Subsystems $A$, $B$, and $C$ for the fractal lattice with $N=81$ sites. (b) Numerical results for $-S_A-S_B-S_C+S_{A\cup B}+S_{B\cup C}+S_{C\cup A}-S_{A\cup B\cup C}$ for $N=81$ and different values of $M$ and $q$. The topological entanglement entropies for the Laughlin states in two dimensions, $\gamma=\ln(2)/2\approx0.3466$ for $q=2$ and $\gamma=\ln(3)/2\approx0.5493$ for $q=3$, are shown as horizontal lines. As explained in the text, we do not expect the numerical values to agree with the topological entanglement entropies for the Laughlin states, since the states on the fractal do not follow the area law. (c-d) The same as (a-b), but for a fractal with $N=243$ sites. The conclusion is unchanged.}\label{fig:TEE}
\end{figure}

Since we observed above that the entropy of the Laughlin states on the fractal lattice does not follow the area law, we are not able to compute the topological entanglement entropy by using the Kitaev-Preskill or the Levin-Wen methods. We nevertheless do the computation to see how far the results are from the topological entanglement entropy $\ln(q)/2$ for the Laughlin states. The Kitaev-Preskill method \cite{kitaev2006} considers three subsystems $A$, $B$, and $C$ such as those shown in Fig.\ \ref{fig:TEE}(a,c). Each subsystem should be large compared to the correlation length, which is the case here. One then computes $-S_A-S_B-S_C+S_{A\cup B}+S_{B\cup C}+S_{C\cup A}-S_{A\cup B\cup C}$, which converges to the topological entanglement entropy $\gamma$ in the limit of large subsystem sizes if the system fulfils the area law.

Our results for this combination of entanglement entropies are shown in Fig.\ \ref{fig:TEE}(b,d), where we also show the topological entanglement entropy for the Laughlin states in two dimensions as horizontal lines. As expected, there are large discrepancies, and the discrepancies are seen both for the lattice with $81$ sites and for the lattice with $243$ sites. In contrast, we find that the numerical value of $\gamma$ computed for a square lattice at half filling with $16\times 16$ sites, employing the same approach, matches the expected value. It remains an interesting open question, whether one can find different methods to extract the topological entanglement entropy from entropies computed on the fractal lattice.

\section{Entanglement spectrum}\label{sec:ES}

In \cite{li2008entanglement}, Li and Haldane postulated and demonstrated numerically the use of the low-lying part of the \textit{entanglement spectrum} as a fingerprint of topological order in fractional quantum Hall systems. Although it has turned out to sometimes be difficult to interpret the results \cite{chandran2014}, there are several examples where the entanglement spectrum shows interesting structures \cite{lauchli2010, dubail2012real,sterdyniak2012,dubail2012edge,regnault2017entanglement}.

The entanglement spectrum can be obtained from the Schmidt decomposition as follows. If the system is in a state $|\psi\rangle$, and we can write the Hilbert space $\mathcal{H}$ as $\mathcal{H}_{A} \otimes \mathcal{H}_{B}$, where $A$ and $B$ form a bipartition of the system, we can perform a Schmidt decomposition
\begin{equation}\label{schmidt}
|\psi\rangle\ =
\sum_j e^{-(1/2)\xi_j} |\psi^j_A\rangle \otimes |\psi^j_B \rangle,
\end{equation}
where $|\psi^j_A\rangle$ and $|\psi^j_B \rangle$ are orthonormal bases in $\mathcal{H}_{A}$ and $\mathcal{H}_{B}$, respectively. We have $\exp(-\frac{1}{2} \xi_j) \ge 0$. The $\xi_j$ are referred to as the entanglement spectrum. The entanglement spectrum is a collection of numbers as opposed to the entanglement entropy, which is a single number, and so potentially contains more information about the state.

In this section, we study the entanglement spectrum of the Laughlin state on the fractal lattice and compare it to the entanglement spectrum of the Laughlin state on two-dimensional square and triangular lattices. We find that the number of states in the low-lying part of the entanglement spectrum is consistent across all these lattices and for different bipartitions.

As the Laughlin state contains a well-defined number of particles and has a well-defined angular momentum for lattices with rotational symmetry, we can group the entanglement eigenvalues based on $M_A$ and $L_{z,A}$, where $M_A$ is the number of particles in part $A$ and $L_{z,A}$ is the angular momentum of part $A$. If we only group with respect to $M_A$, we can compare the entanglement spectra across all three kinds of lattices and consider various bipartitions, as long as the sizes of both subsystems $A$ and $B$ are large enough. When we group the entanglement eigenvalues with respect to both $M_A$ and $L_{z,A}$, however, we can only compare the fractal and triangular lattices, since only these two lattices share a $C_{3}$ rotational symmetry about their centers. Also, one can only consider subsystems, which have the same rotational symmetry as the whole lattice in addition to the constraints mentioned above.

We also do the computation for a randomly chosen state. For this state we find that the number of states below the entanglement gap is much larger than for the Laughlin state, and the number of states below the entanglement gap changes with the choice of lattice and bipartition. The consistency of the number of states below the entanglement gap for the Laughlin state and a different behavior for a randomly chosen state is the behavior expected from the Li-Haldane conjecture. We hence conclude that the topology of the Laughlin states on the fractal lattice can be seen in the entanglement spectrum.

\subsection{Results}

\begin{figure}
\includegraphics[width=0.9\linewidth,trim=5mm 85mm 20mm 85mm,clip]{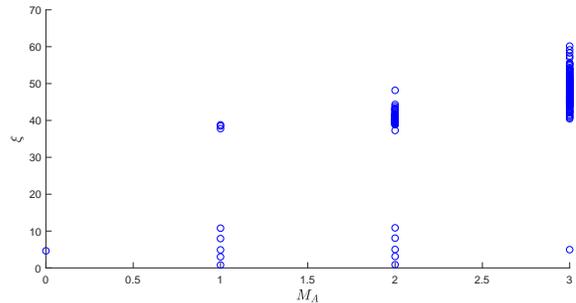}
\caption{Entanglement spectrum for the fractal lattice with $N=27$ sites and $M=3$ particles. The selected region consists of the 12 sites marked in Fig.~\ref{fig:fractal_paper}. The counting sequence $1,5,5,1$ of eigenvalues below the gap is robust with respect to the choice of bipartition and to the nature of the lattice.}
\label{fig:ES_m_frac}
\end{figure}

Figure \ref{fig:ES_m_frac} gives an example of the spectra we obtain when $N_{\phi}=qM$ is less than about $16$ and $N$ is around $27$. We see an entanglement gap separating a low-lying, discrete part from the higher, more continuous part. While the number of states in the higher part changes with the choice of bipartition, the number of states below the gap is robust. If we change the lattice from a fractal to a two-dimensional triangular or square lattice (which requires changing the total number of sites $N$) while keeping $q$ and $M$ constant, the number of states below the gap still remains the same.

The results for the number of states below the entanglement gap for different values of $q$ and $M$ are listed in Tab.\ \ref{tab:count}. The counting in the $M_{A}$ sectors is given in the second to last column. From left to right, the numbers denote the number of states below the gap, starting from $M_{A}=0$ to $M_{A}=M$. The last column indicates how each number in the preceding column is divided into the three possible $L_{z,A}$ values $(0,2\pi/3,4\pi/3)$. For $q=2$ and $M=3$, e.g., the counting of states below the gap in the $M_{A}$ sectors is $1\hspace{0.2cm}5\hspace{0.2cm} 5\hspace{0.2cm} 1$, as shown in Fig.~\ref{fig:ES_m_frac}. The first state, corresponding to $M_{A}=0$ has an angular momentum of $0$. The five states in the $M_{A}=1$ sector consist of one state with $L_{z,A}=0$, three states with $L_{z,A}=2\pi/3$, and one state with $L_{z,A}=4\pi/3$.

We have also computed the entanglement spectrum for a state with randomly chosen coefficients but all other parameters and the lattice kept the same. In this case, the number of states below the gap is different and varies with respect to the chosen bipartition. This points to the conclusion that the counting we have obtained is a salient feature of the Laughlin state and signifies its topological nature.

\begin{table*}
\begin{tabular}{ccc p{1mm} l p{1mm} l}
\hline
\hline
$q$ & $M$ & $N_{\phi}$ && Counting($M_{A}$) && Counting($M_A$,$L_{z,A}$)\\
\hline
2 & 1 &  2 &&   1 1 && 1 0 0 -- 0 0 1\\
2 & 2 &  4 && 1 3 1 && 1 0 0 -- 1 1 1 -- 0 0 1\\
2 & 3 &  6 &&  1 5 5 1 && 1 0 0 -- 1 3 1 -- 1 3 1 -- 0 1 0 \\
2 & 4 &  8 &&  1 7 15 7 1 && 1 0 0 -- 2 4 1 -- 5 9 1 -- 2 4 1 -- 0 1 0\\
2 & 5 & 10 && 1 9 28 28 9 1 && 1 0 0 -- 3 5 1 -- 9 14 5 -- 9 14 5 -- 3 5 1 -- 0 1 0 \\
2 & 6 & 12 && 1 11 45 84 45 11 1 && 1 0 0 -- 3 7 1 -- 16 18 11 -- 29 31 24 -- 16 18 11 -- 3 7 1 -- 0 1 0\\	
\hline
\hline
\end{tabular}

\vspace{4mm}

\begin{tabular}{ccc p{1mm} l p{1mm} l}
\hline
\hline
$q$ & $M$ & $N_{\phi}$ && Counting($M_{A}$) && Counting($M_A$,$L_{z,A}$)\\
\hline
3 & 1 &  3 &&   1 1 && 1 0 0 -- 0 0 1\\
3 & 2 &  6 && 1 4 1 && 1 0 0 -- 1 2 1 -- 0 0 1\\
3 & 3 &  9 && 1 7 7 1 && 1 0 0 -- 2 4 1 -- 2 4 1 -- 0 1 0\\
3 & 4 & 12 && 1 10 28 10 1 && 1 0 0 -- 3 6 1 -- 9 14 5 -- 3 6 1 -- 0 1 0 \\
\hline
\hline
\end{tabular}

\vspace{4mm}

\begin{tabular}{ccc p{1mm} l p{1mm} l}
\hline
\hline
$q$ & $M$ & $N_{\phi}$ && Counting($M_{A}$) && Counting($M_A$,$L_{z,A}$)\\
\hline
4 & 1 &  4 && 1 1 && 1 0 0 -- 0 0 1\\
4 & 2 &  8 && 1 5 1 && 1 0 0 -- 1 3 1 -- 0 0 1\\
4 & 3 & 12 && 1 9 9 1 && 1 0 0 -- 3 5 1 -- 3 5 1 -- 0 1 0\\
4 & 4 & 16 && 1 13 45 13 1 && 1 0 0 -- 4 8 1 -- 16 18 11 -- 4 8 1 -- 0 1 0\\
\hline
\hline		
\end{tabular}

\caption{The number of states below the entanglement gap in the sectors with fixed $M_{A}$ and fixed $L_{z,A}$. The column labeled Counting($M_{A}$) gives the number of states below the entanglement gap for $M_A=0,1,\ldots,M$, and the column labeled Counting($M_A$,$L_{z,A}$) gives the number of states below the entanglement gap for given $M_A$ and $L_{z,A}$. The dashes separate different $M_A$ values, and for each $M_A$ value there are three possible values of the angular momentum. As an example, the second row of the first table should be read as follows. There is one state below the entanglement gap with $M_A=0$, three states below the entanglement gap with $M_A=1$, and one state below the entanglement gap with $M_A=2$. The state with $M_A=0$ has angular momentum $L_{z,A}=0$, the states with $M_A=1$ have angular momenta $L_{z,A}=0$, $L_{z,A}=2\pi/3$, and $L_{z,A}=4\pi/3$, respectively, and the state with $M_A=2$ has angular momentum $L_{z,A}=4\pi/3$.}\label{tab:count}
\end{table*}

\begin{figure}
\includegraphics[width=0.5\linewidth,trim=64mm 92mm 59mm 88mm,clip]{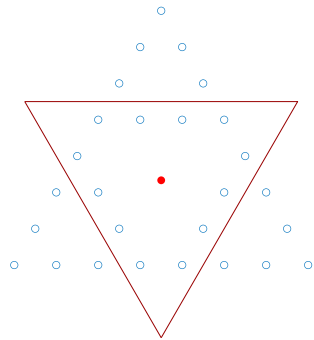}
\caption{The region of the fractal lattice used for computing the entanglement spectrum as a function of angular momentum consists of the sites inside the triangle. Both the complete fractal lattice and the two subsystems have a three-fold rotation symmetry around the point indicated by the red dot.}
\label{fig:fractal_paper}
\end{figure}

\section{Conclusion}\label{sec:conclusion}

We have studied density, correlation, and entanglement properties of Laughlin states on fractal lattices derived from the Sierpinski triangle and identified several similarities and differences compared to the corresponding properties of Laughlin states on two-dimensional lattices. Laughlin states on two-dimensional lattices have a uniform density in the bulk, exponentially decaying particle-particle correlation functions, area law entanglement entropy, and a particular counting of states below the entanglement gap. On the fractal lattices, there is more structure in the density patterns, although the density still stays roughly uniform, except for the corner sites. The particle-particle correlation functions also decay roughly exponentially on the fractal lattice. The entanglement entropy, however, does not follow the area law. Instead we find that the scaling of the entanglement entropy with the subsystem size is close to logarithmic for subsystems with similar boundaries. This also means that the topological entanglement entropy cannot be obtained from the Kitaev-Preskill or the Levin-Wen approach. In addition, we observe oscillations of the entropy as a function of $M$ that are not observed for Laughlin states on two-dimensional square lattices. Finally, we find that the number of states below the entanglement gap is the same for the fractal and the two-dimensional lattices. We have here considered Laughlin trial states because they allow us to investigate particularly large systems and hence obtain a clearer picture, but we expect similar results to hold more broadly.

A number of tools have been developed to detect topological properties of quantum many-body systems defined on two-dimensional lattices. The results presented here provide information about which of these tools can be used for fractal lattices and which cannot. Demonstrating the ability of the states to host anyons is a direct way to detect topology, and this approach was used already in \cite{manna2019} to confirm the topology of the Laughlin states on fractal lattices. On the other hand, Chern number and spectral flow do not translate naturally to fractals, as the fractal lattices do not have periodic boundary conditions. The present work, which considers fractals derived from the Sierpinski triangle, shows that the entanglement spectrum can be used to probe topology for such fractals, while standard methods to extract the topological entanglement entropy that rely on the area law fail for such fractals. We expect the former conclusion to hold more generally, while the latter conclusion may depend on, e.g., the dimension or ramification of the fractal lattice. The fewer tools available to detect topology in systems defined on fractals motivates the search for further tools and poses the challenge to find out whether known methods that do not immediately work for fractals can be modified in a way that nevertheless make them applicable to fractals.

A key motivation for constructing fractional quantum Hall type models on fractal lattices is to find systems with combinations of properties that are not seen elsewhere. The results obtained here indeed show that the Laughlin states on fractal lattices have important differences compared to Laughlin states on two-dimensional lattices. They do, e.g., provide the opportunity to obtain fractional quantum Hall physics in systems without area law entanglement. This encourages the search for further unusual properties of systems on fractal lattices.

\begin{acknowledgments}
This work has been supported by the Independent Research Fund Denmark under grant number 8049-00074B and the Carlsberg Foundation under grant number CF20-0658.
\end{acknowledgments}

\end{document}